\documentclass[%
 reprint,
superscriptaddress,
amsmath,amssymb,
prl,
floatfix,
]{revtex4-1}
\usepackage{breqn}
\usepackage{graphicx}
\usepackage{dcolumn}
\usepackage{bm}
\usepackage{verbatim} 
\usepackage{fancyhdr}

\usepackage[
  pdfusetitle,
  bookmarks,bookmarksopen,
  bookmarksnumbered,
  pdfpagelabels,
  colorlinks,linkcolor=blue,citecolor=red,urlcolor=blue,
  pdfview={Fit},
]{hyperref}
\usepackage[mathlines]{lineno}

\newcommand{\numu}{\nu_{\mu}}
\newcommand{\nue}{\nu_{e}}
\newcommand{\pio}{\pi^{0}}

\newcommand{\cm}{\text{cm}}
\newcommand{\ar}{\text{Ar}}

\begin{document}

\title{First Measurement of $\numu$ Charged-Current $\pio$ Production on Argon with a LArTPC}

%
\newcommand{\Bern}{Universit{\"a}t Bern, Bern CH-3012, Switzerland}
\newcommand{\BNL}{Brookhaven National Laboratory (BNL), Upton, NY, 11973, USA}
\newcommand{\Cambridge}{University of Cambridge, Cambridge CB3 0HE, United Kingdom}
\newcommand{\Chicago}{University of Chicago, Chicago, IL, 60637, USA}
\newcommand{\Cincinnati}{University of Cincinnati, Cincinnati, OH, 45221, USA}
\newcommand{\CSU}{Colorado State University, Fort Collins, CO, 80523, USA}
\newcommand{\Columbia}{Columbia University, New York, NY, 10027, USA}
\newcommand{\Davidson}{Davidson College, Davidson, NC, 28035, USA}
\newcommand{\FNAL}{Fermi National Accelerator Laboratory (FNAL), Batavia, IL 60510, USA}
\newcommand{\Harvard}{Harvard University, Cambridge, MA 02138, USA}
\newcommand{\IIT}{Illinois Institute of Technology (IIT), Chicago, IL 60616, USA}
\newcommand{\KSU}{Kansas State University (KSU), Manhattan, KS, 66506, USA}
\newcommand{\Lancaster}{Lancaster University, Lancaster LA1 4YW, United Kingdom}
\newcommand{\LANL}{Los Alamos National Laboratory (LANL), Los Alamos, NM, 87545, USA}
\newcommand{\Manchester}{The University of Manchester, Manchester M13 9PL, United Kingdom}
\newcommand{\MIT}{Massachusetts Institute of Technology (MIT), Cambridge, MA, 02139, USA}
\newcommand{\Michigan}{University of Michigan, Ann Arbor, MI, 48109, USA}
\newcommand{\NMSU}{New Mexico State University (NMSU), Las Cruces, NM, 88003, USA}
\newcommand{\Otterbein}{Otterbein University, Westerville, OH, 43081, USA}
\newcommand{\Oxford}{University of Oxford, Oxford OX1 3RH, United Kingdom}
\newcommand{\PNNL}{Pacific Northwest National Laboratory (PNNL), Richland, WA, 99352, USA}
\newcommand{\Pitt}{University of Pittsburgh, Pittsburgh, PA, 15260, USA}
\newcommand{\StMarys}{Saint Mary's University of Minnesota, Winona, MN, 55987, USA}
\newcommand{\SLAC}{SLAC National Accelerator Laboratory, Menlo Park, CA, 94025, USA}
\newcommand{\Syracuse}{Syracuse University, Syracuse, NY, 13244, USA}
\newcommand{\TelAviv}{Tel Aviv University, Tel Aviv, Israel, 69978}
\newcommand{\Tennessee}{University of Tennessee, Knoxville, TN, 37996, USA}
\newcommand{\UTA}{University of Texas, Arlington, TX, 76019, USA}
\newcommand{\Tubitak}{TUBITAK Space Technologies Research Institute, METU Campus, TR-06800, Ankara, Turkey}
\newcommand{\Tufts}{Tufts University, Medford, MA, 02155, USA}
\newcommand{\VTech}{Center for Neutrino Physics, Virginia Tech, Blacksburg, VA, 24061, USA}
\newcommand{\Yale}{Yale University, New Haven, CT, 06520, USA}

\affiliation{\Bern}
\affiliation{\BNL}
\affiliation{\Cambridge}
\affiliation{\Chicago}
\affiliation{\Cincinnati}
\affiliation{\CSU}
\affiliation{\Columbia}
\affiliation{\Davidson}
\affiliation{\FNAL}
\affiliation{\Harvard}
\affiliation{\IIT}
\affiliation{\KSU}
\affiliation{\Lancaster}
\affiliation{\LANL}
\affiliation{\Manchester}
\affiliation{\MIT}
\affiliation{\Michigan}
\affiliation{\NMSU}
\affiliation{\Otterbein}
\affiliation{\Oxford}
\affiliation{\PNNL}
\affiliation{\Pitt}
\affiliation{\StMarys}
\affiliation{\SLAC}
\affiliation{\Syracuse}
\affiliation{\TelAviv}
\affiliation{\Tennessee}
\affiliation{\UTA}
\affiliation{\Tubitak}
\affiliation{\Tufts}
\affiliation{\VTech}
\affiliation{\Yale}

\author{C.~Adams} \affiliation{\Yale}
\author{M.~Alrashed} \affiliation{\KSU}
\author{R.~An} \affiliation{\IIT}
\author{J.~Anthony} \affiliation{\Cambridge}
\author{J.~Asaadi} \affiliation{\UTA}
\author{A.~Ashkenazi} \affiliation{\MIT}
\author{M.~Auger} \affiliation{\Bern}
\author{S.~Balasubramanian} \affiliation{\Yale}
\author{B.~Baller} \affiliation{\FNAL}
\author{C.~Barnes} \affiliation{\Michigan}
% ...

\collaboration{The MicroBooNE Collaboration\footnote{microboone\_info@fnal.gov}} \noaffiliation

\date{\today}

\begin{abstract}
We report the first measurement of the flux-integrated cross section of $\numu$~charged-current single $\pio$~production on argon. This measurement is performed with the MicroBooNE detector, an 85~ton active mass liquid argon time projection chamber exposed to the Booster Neutrino Beam at Fermilab. This result on argon is compared to past measurements on lighter nuclei to investigate the scaling assumptions used in models of the production and transport of pions in neutrino-nucleus scattering. The techniques used are an important demonstration of the successful reconstruction and analysis of neutrino interactions producing electromagnetic final states using a liquid argon time projection chamber operating at the earth's surface.
\begin{description}
\item[PACS numbers] 13.15.+g,13.60.Le,25.30.Pt
\end{description}
\end{abstract}

\maketitle

Neutral pion ($\pio$) production in neutrino interactions can create backgrounds that limit the sensitivity of neutrino $\numu\rightarrow\nu_{e}$ oscillation searches, such as those being pursued by DUNE~\cite{duneIDR1,duneIDR2,duneIDR3} and the SBN Program~\cite{SBN}. This background comes from photons originating from the $\pio\rightarrow\gamma\gamma$ decays that can mimic the topology of an electron originating from a $\nue$ charged-current interaction. Uncertainties associated with this background can have a detrimental impact on experimental searches for the appearance of $\nue$ in $\numu$-beams.

To reduce this uncertainty, the physics underlying the primary interaction and the subsequent transport of hadrons through the nuclear medium needs to be understood across the wide range of targets used in neutrino experiments. While $\numu$ charged-current (CC) with an associated $\pio$ has been previously studied on deuterium by bubble chamber experiments~\cite{bub_anl,bub_bnl1,bub_bnl2,BEBC}, on freon~\cite{SKAT}, and on carbon~\cite{minpi0_1,minpi0_2,mbpi0,KEK}, it has never been measured on heavier targets. This letter reports the first measurement of $\numu$~CC single $\pio$~production on argon using an inclusive final-state topology of at least one photon coming from a $\pio$ decay in addition to a muon, both exiting the target nucleus. 

{\it Experimental Setup and Simulation $-$} This measurement is performed using neutrinos originating from the Booster Neutrino Beam (BNB) at Fermilab~\cite{bnb}. The BNB creates a 93.6\% pure source of $\numu$. Sitting 463~m from the target, the MicroBooNE detector is a LArTPC with 85~tons of active mass~\cite{uBdet} that is read out by three planes of sense wires. As charged particles traverse the argon, the resulting ionization electrons drift in an electric field to the wires. The first two sense planes record bipolar signals while the final sense plane collects the charge providing a unipolar signal, which produces a measure of the deposited energy. Scintillation light produced during the ionization process is collected by an array of photomultiplier tubes (PMTs). A readout is selected for further analysis by a coincidence of PMT light in a beam spill. This analysis makes use of a data sample corresponding to $1.62\times10^{20}$ protons on target, after passing data and beam quality requirements, collected between February and July 2016.

The flux of neutrinos at MicroBooNE is simulated using the framework built by the MiniBooNE collaboration~\cite{bnb}. Neutrino interactions on argon, along with the relevant nuclear processes that modify the final state, are simulated with the GENIE event generator~\cite{genie}, which is commonly used in many neutrino experiments. We configure GENIE with a relativistic Fermi gas nuclear model~\cite{fermigas} with additional terms that enhance quasi-elastic-like interactions that occur off of correlated nucleon pairs via an empirically driven meson exchange current model~\cite{mec}. Resonant pion production is described by the Rein-Sehgal (RS) model~\cite{rs} that scales the cross section based on the number of neutrons in a nuclear target. The effects of final state interactions (FSI) are handled by an effective cascade model ({\it hA})~\cite{genieuncert} which scales the hadron-nucleus cross sections as $A^{2/3}$. 

Particles exiting the incident nucleus are passed to a custom implementation of {\sc Geant4} available in the LArSoft software toolkit~\cite{g4,larsoft}. Backgrounds induced by cosmic rays (CRs) that produce activity coinciding with the beam spill are measured directly in data by utilizing a random trigger anti-coincident with the beam. Other CR backgrounds are modeled with CORSIKA~\cite{corsika} at an elevation of $226$~m above sea level.

{\it Reconstruction and Event Selection $-$ } Signal processing begins by filtering electronics noise from the raw wire signals~\cite{noise}. The filtered waveforms are processed to isolate Gaussian shaped signals~\cite{hits}, called hits. The Pandora event reconstruction toolkit~\cite{pandora} is used to cluster the hits and construct 3D tracks and vertices. The 3D vertices are candidate locations for neutrino interactions and are used in the next stage of the reconstruction as a seed for shower finding.

To remove CR muons, tracks that cross any two detector faces are removed. Tracks that are inconsistent with the spatial distribution of light detected in the PMT array are also removed. The remaining tracks are treated as candidate neutrino-induced muons.

A candidate muon track must have a length greater than 15~cm and be matched to within 3~cm of a three-dimensional (3D) reconstructed vertex that is located within a fiducial volume taken as 10~cm from the upstream and downstream faces of the detector and 20~cm from the sides. Any additional tracks that have one endpoint within 3~cm of this same vertex are considered associated with the vertex. 
To further reduce CR contamination, a set of multiplicity-dependent requirements are applied. Single detector-exiting tracks, vertical tracks, two-track topologies compatible with a muon decaying, and multi-track vertices where the two longest tracks are back-to-back ($>155^{\circ}$) are rejected. Finally, we require that the candidate muon track has a deposited charge consistent with a minimally ionizing particle and has no deflection at any point along the track greater than 8$^{\circ}$; this removes tracks mistakenly reconstructed from EM particle showers.

The number of preselected CR events, as measured in data, are reduced by 99.9\%. From the simulation we find the remaining sample consists of 80\% $\numu$ CC interactions (6\% with a single $\pio$) and 15\% CR backgrounds, with the remainder being neutral-current and CC interactions from other neutrino flavors. The efficiency for selecting a CC single $\pio$ event is 33\% based solely on finding the muon track. To identify the $\numu$ CC single $\pio$ events, a second pass reconstruction is employed to identify electromagnetic (EM) particles associated with the neutrino interaction.

This reconstruction constitutes the first demonstration of EM particle 3D reconstruction without human assistance in LArTPC data. 
The reconstruction of EM activity (EM showers) is separated into two stages: the first aims to sort hits to identify neutrino-induced EM activity, and the second clusters these hits into individual showers. The first stage begins by seeding the EM shower reconstruction on each readout plane with the output of the earlier clustering pass performed by Pandora. This Pandora clustering pass collects charge fragments from single parent particles but spreads the full charge of that particle across many clusters~\cite{pandora}. 
If the cluster appears to not be shower-like (based on its linearity), emanates from a track-like particle, or is uncorrelated with the interaction vertex, it is rejected~\cite{dc}. This procedure becomes inefficient for particles with kinetic energies below 50~MeV, as radiative and ionization energy losses become comparable, resulting in linear trajectories for EM particles~\cite{michel,dc}. In the second stage of EM shower reconstruction, the hits designated as ``shower-like'' are passed to a  re-clustering procedure  using OpenCV, an open source image processing tool~\cite{opencv,ahack}, that processes the hits radially from the candidate neutrino vertex. During image processing, all contiguous hits are formed into a 2D cluster on a given plane. The resulting OpenCV clusters are matched across planes by matching the time ranges of the clusters on the collection plane with one of the two induction planes. With matched clusters, shower properties such as 3D direction and energy from the summed hit charge on the collection plane are calculated. 

\begin{figure}
\centering
\includegraphics[width=1\columnwidth]{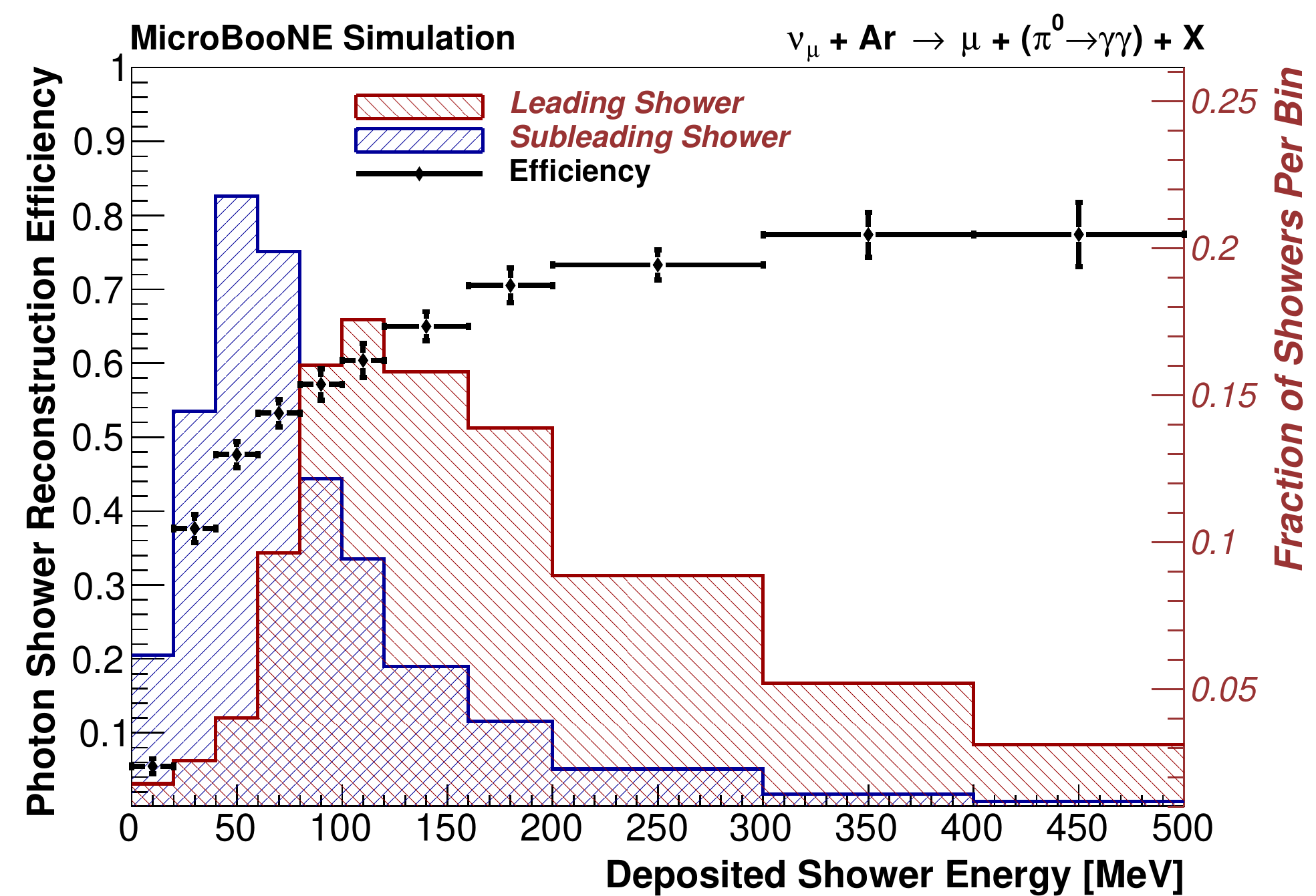}
\caption{ The $\numu+\text{Ar}\rightarrow\mu+(\pio\rightarrow\gamma\gamma)+X$ shower reconstruction efficiency as a function of the deposited energy of the shower (with statistical uncertainties). Overlaid are the simulated energy distributions of the decay photons from neutrino induced neutral pions. The leading shower spectrum is in red and the subleading shower in blue.}
\label{fig:eff}
\end{figure}

The algorithm results in relatively high purity showers (on average 92\% of the collected charge comes from the same parent particle) at the expense of completeness (on average, a reconstructed shower contains 63\% of the total charge associated to the parent particle). To estimate the shower reconstruction efficiency we start with $\numu+~\text{Ar}\rightarrow\mu+\pio+X$ interactions that pass our preselection. We then apply our two stage shower reconstruction and plot the ratio of the number of reconstructed showers over all showers as a function of true deposited photon energy, as shown in Fig.~\ref{fig:eff}, along with the simulated leading and subleading photon deposited energy distributions. Less than 1\% of photons deposit energy within the detector but result in no identifiable charge on the collection plane. The low efficiencies at low energies are instead due to the removal of track-like topologies to mitigate CR contamination for this charged-current $\numu$ analysis. More sophisticated techniques for identifying EM showers, like deep neural-networks~\cite{ssnet}, could be employed in the future to increase our efficiencies at the lowest energy. 

At BNB neutrino energies more than $95\%$ of neutrino induced photons are 
predicted to come from single $\pio$ decays, with the remainder predominantly coming from events with two or more $\pio$ decays. For the cross section measurement described here, we require at least one photon to be reconstructed, enabling a higher event selection efficiency. The efficiency and background subtraction used are estimated from the simulation and a two shower selection is employed as a cross check. To associate a shower to the neutrino interaction, we require at least one reconstructed shower to point towards the interaction vertex with an impact parameter, or distance of closest approach of the backward shower projection, of less than 4~cm, and a start point located within $62$~cm of the vertex. These values are chosen to maximize the purity of the selection.

Requiring one or more reconstructed photons, there are 771~candidate events in the data sample that, based on simulation, has a 56\% purity and 16\% efficiency for $\numu$ CC $\pio$ interactions. The dominant source of background, 15\% of the sample, comes from real EM showers produced near the vertex (such as radiation emanating from muons), Michel decays, $\pi^{\pm}\rightarrow\pio$ charge-exchange in pion transit in the detector, and nucleon inelastic scatters in the detector volume. A further 8\% of the events have a shower misreconstructed from non-EM activity. CR backgrounds make up 12\% of the sample. The remaining sample results from multi-$\pio$ events (5\%), $\numu$ CC induced single $\pio$ events outside the fiducial volume (2\%), and the remainder come from neutral current and non-$\numu$ CC interactions.

\begin{figure}
\centering
\includegraphics[width=1\columnwidth]{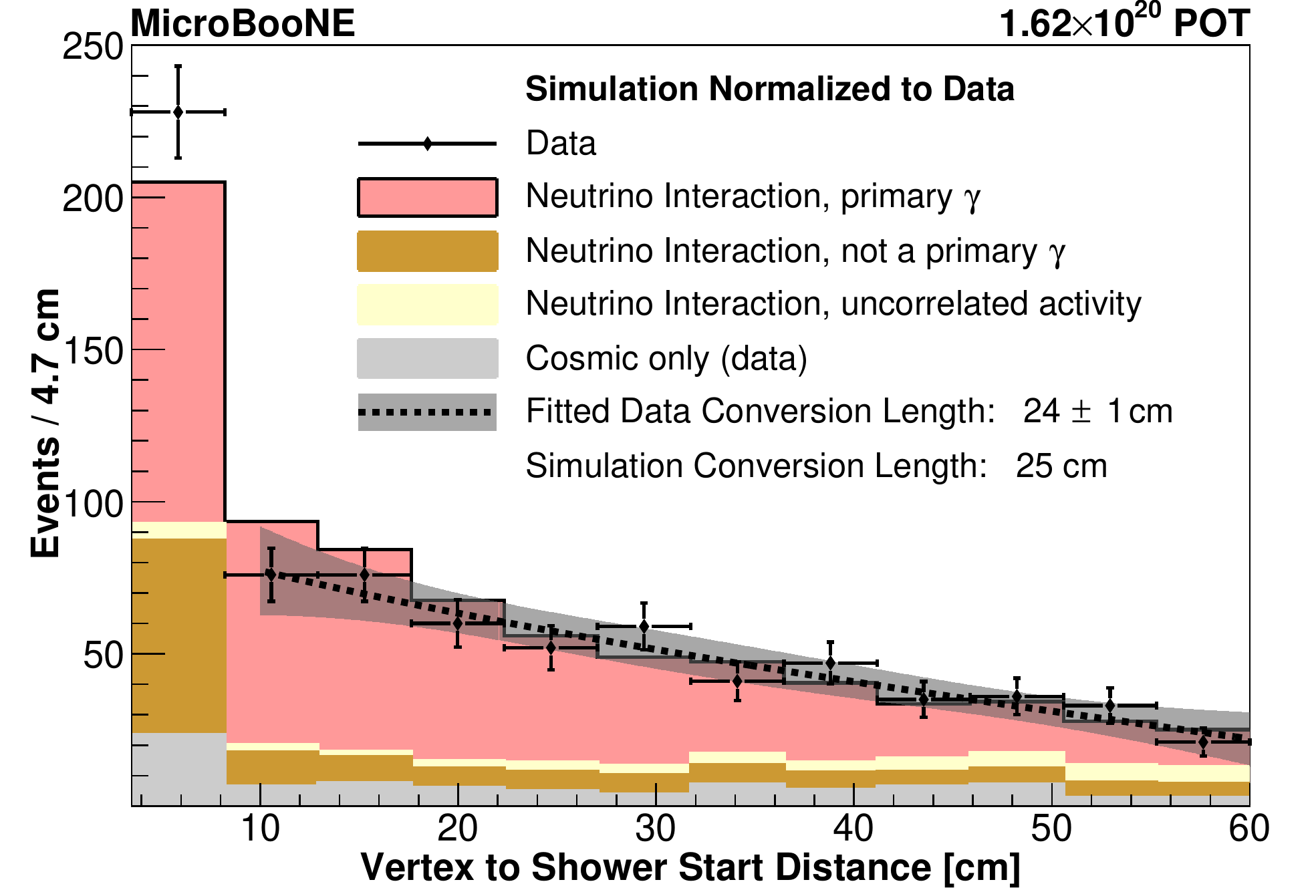}
\caption{The vertex to shower start point distance in events with at least one shower correlated with the neutrino interaction vertex. The histogram of simulated neutrino interactions have been area normalized to the data to enable a shape comparison and separated into four classes: selected neutrino-induced primary photons (red), selected activity from the neutrino interaction which is not a primary photon (brown), activity uncorrelated with the neutrino interaction (yellow), and pure CR backgrounds (gray). The fit for these backgrounds and the extracted conversion length excludes the first bin.}
\label{fig:cd}
\end{figure}

The distribution of the 3D distance from a vertex to the reconstructed shower start point is shown in Fig.~\ref{fig:cd} along with a breakdown of the selected sample into primary photons created by: a neutrino interaction, activity from a neutrino interaction we identify as a shower that is not a primary photon, activity uncorrelated with the neutrino interaction (noise or CR) misidentified as a shower coming from the neutrino interaction vertex, and purely CR induced backgrounds, where the simulation is area normalized to the data. This distribution is fit, in the range of 13~cm to 60~cm, with an exponential function whose slope provides a measurement of the conversion distance of the photons. We exclude the first bin from the fit to remove the contribution from tracks misreconstructed as showers near the vertex.  A linear function is included in the fit to model the summed backgrounds, which tend to be flat based on simulation. The resulting conversion distance of $24\pm 1~(\text{stat})$~cm is consistent with simulation and consistent with our expectation of the energy dependent photon-argon cross section~\cite{dc}. 

To cross-check this selection, we measure the two-photon invariant mass spectrum with a second selection that requires at least two showers reconstructed with an impact parameter less than 4~cm. The leading photon of a $\pio$ decay cannot have less energy than  $m_{\pio}/2$, therefore, it is required that at least 60\% of the photon energy is reconstructed (40~MeV). Reconstructed showers that are separated by less than $20^{\circ}$ are largely the result of a single photon being reconstructed as two showers and are rejected. Finally, the leading and subleading showers are required to start within 80~cm and 100~cm of the interaction vertex, respectively. Events where more than one pair of showers pass this criteria are rejected as multi-$\pio$ background. This two-shower selection has a purity of 64\% and a signal efficiency of 6\%, based on simulation. 

\begin{figure}
\centering
\includegraphics[width=1\columnwidth]{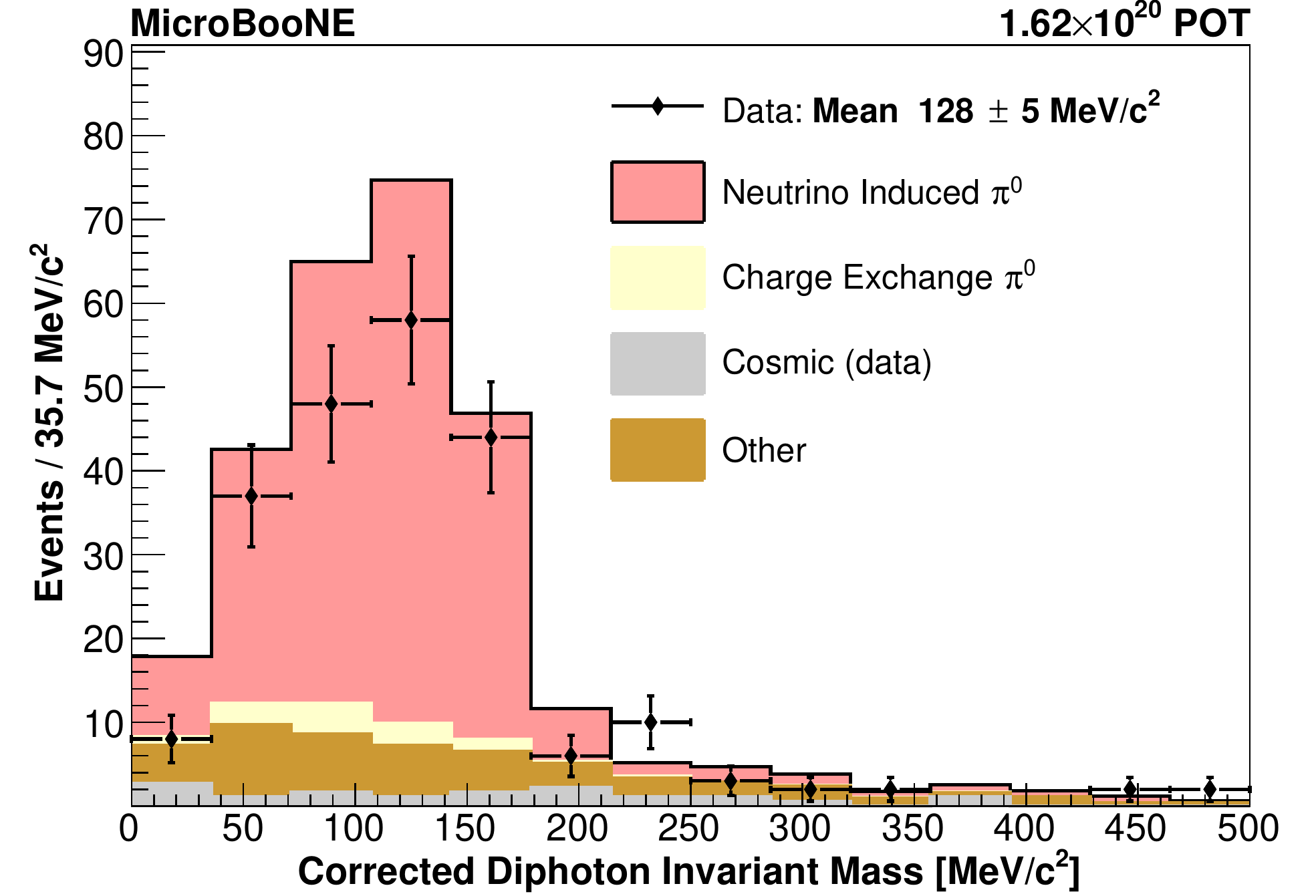}
\caption{ The reconstructed invariant mass of the two photon candidates associated to the neutrino interaction vertex after an energy-scale correction. The simulation is grouped into four classes of photon pairs: neutrino-induced $\pio$ that are created in and subsequently exit the argon nucleus (red), charged pion charge exchange that occurs outside the nucleus (yellow), pure CR activity (grey), and everything else (brown). The mean of the data is consistent, within the quoted statistical uncertainty, with  $m_{\pi^{0}}$.}
\label{fig:m}
\end{figure}

With two showers, the diphoton invariant mass is measured and compared with the expected $\pio$ mass. We apply simulation-based shower energy-scale corrections of, on average, 40\% to account for energy lost during hit formation and clustering~\cite{dc}. The final diphoton invariant mass distribution has a mean, $128\pm5~\text{MeV}/\text{cm}^2$, consistent, within statistical uncertainties, with the $\pio$ mass (Fig.~\ref{fig:m}). The normalization disagreement shown in Fig.~\ref{fig:m} is within flux and cross section uncertainties, discussed later. This provides further confidence that the selected photons originate from $\pio$ decays.

{\it Cross Section Measurement $-$ } Using the selection with at least one-shower, we measure the total flux integrated cross section via the following relation:
\begin{equation}
\left\langle\sigma\right\rangle_{\Phi} = \frac{N - B}{\epsilon  T \Phi},
\end{equation}
\noindent where $N$ is the number of events selected in data (771~events), $B$ is the number of expected background events, $\epsilon$ is the efficiency for selecting signal events, $T$ is the number of argon targets within the fiducial volume, and $\Phi$ is the integrated $\numu$ flux from 0~GeV to 3~GeV. Off-beam data are used to model the pure CR backgrounds in $B$ (86.9~events); the remainder of the total background (347.3~events) are taken from the simulation. The detector volume is treated as pure argon to calculate $T$.

We identify three major sources of systematic uncertainty for this measurement: the neutrino flux prediction, the neutrino-argon interaction model, and the detector simulation. We assess uncertainties on the neutrino flux prediction using the final flux simulation from the MiniBooNE collaboration~\cite{bnb} adopted to the MicroBooNE detector size and location. These account for hadron production in the beamline, the focusing optics of the secondary pion beam, and proton counting. Varying these effects results in a 16\% uncertainty on the final cross section. For the neutrino-argon interaction uncertainties, individual parameters are varied within the GENIE neutrino interaction models~\cite{genieuncert}. The dominant uncertainties on the backgrounds come from the resonance model parameterization and the FSI modeling and lead to a 17\% total uncertainty on the resulting cross section measurement.  Finally, for the detector simulation, a wide variety of microphysical effects are varied, including the electron diffusion, the scintillation light yield, the electron recombination~\cite{recomb}, and localized electric field distortions. Further, the simulated detector response is varied for effects such as the single photon rate observed in the PMTs, the electronics noise~\cite{noise}, the signal response shape, non-responsive channels, the visibility of the region surrounding the TPC to the PMT array, and a simulation of long-range induced signals on the wires~\cite{signal,signal2}. An additional uncertainty is assessed on the reconstructed neutrino interactions that are contaminated by simulated CR activity. Together the detector simulation variations yield a 21\% uncertainty on the final cross section measurement. This set of uncertainties, while dominant, are expected to be reduced by an ongoing program of detector calibrations.  Each systematic uncertainty is treated as uncorrelated and quadratically summed to give a total systematic uncertainty of 31\%. 

\begin{figure}
\centering
\includegraphics[width=1\columnwidth,trim={0.5cm 6.6cm 0.15cm 6.5cm},clip]{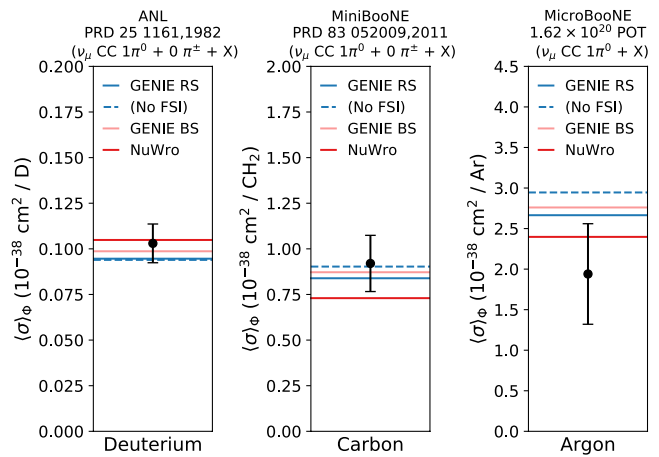}
\caption{ 
The measured total flux integrated $\numu$ CC single $\pio$ cross section for ANL, MiniBooNE, and MicroBooNE with the bars denoting the total uncertainty. These are compared to the flux averaged default GENIE prediction with the RS model (solid blue) and with FSI removed (dashed blue) and an alternative GENIE model with the BS model (solid pink). {\sc NuWro} predictions are shown in solid red. }
\label{fig:xsec}
\end{figure}

{\it Results and Discussion $-$ } The flux-integrated total cross section for CC single $\pio$ production on argon, measured through the reconstruction of at least one shower, is found to be
\begin{equation*}
\left\langle\sigma\right\rangle_{\Phi}=1.9\pm0.2~(\text{stat})\pm0.6~(\text{syst})\times10^{-38}~\frac{\cm^2}{\ar}.
\end{equation*}
\noindent Using the selection that requires at least two showers a consistent cross section, within statistical uncertainties, is measured. We compare four models of resonant pion production to this measurement in Fig.~\ref{fig:xsec}. The RS model~\cite{rs}, shown with and without the effects of FSI, and the Berger-Sehgal (BS) model~\cite{bs}, as implemented in GENIE; as well as for an alternative generator, {\sc NuWro}~\cite{nuwro}. {\sc NuWro} utilizes a local Fermi gas model for the initial nuclear state. Resonant pion production is described via the Adler model~\cite{adler1,adler2} with modified form factors~\cite{nuwroFF}, and the Oset model~\cite{oset} handles the FSI of the hadrons exiting the struck nucleus.  

The predicted cross section from GENIE includes non-resonant components of 24\% (30\%) for final states that exclude (include) additional charged mesons.  These components will not change between different GENIE models and are modeled differently in {\sc NuWro}. Each model depends on scalings that encapsulate the dependence of the production and FSI across a large range of nuclei. To test these scaling assumptions, we bring together measurements of CC single $\pio$ production performed on other nuclei using similar neutrino energy ranges, including those from the ANL bubble chamber~\cite{bub_anl} and MiniBooNE~\cite{mbpi0}. While the present work includes events with any particles beyond the
single $\pio$ and muon, the MiniBooNE and ANL measurements excluded events with additional charged-mesons. The published neutrino fluxes~\cite{bnb,anl_flux,nuis} have been used to derive flux averaged cross section prediction and the results from deuterium, carbon, and argon are shown together in Fig.~\ref{fig:xsec}. 
Each experiment integrates their unique flux across different energy ranges, which complicates the ability to directly scale between nuclear targets. The model comparisons in Fig.~\ref{fig:xsec} employ the same flux integrations as used to treat the data.
The ANL and MiniBooNE measurements agree with both of the GENIE predictions, whereas a slight deficit ($1.2\sigma$) relative to these models is seen for argon. The measured cross sections on deuterium and argon are in agreement with the {\sc NuWro} predictions, while the measurement on carbon sits 1.2$\sigma$ high. This indicates that the scalings implemented in these models are applicable, within the uncertainties, for neutrino-argon scattering. 

In conclusion, this letter reports the first measurement of CC production of single neutral pions in neutrino-argon scattering. The measurement is compared to a set of models implemented in the GENIE and {\sc NuWro} neutrino event generators, which, based on previous measurements, describe this process well on lighter nuclei. We also find consistency for argon. This analysis makes use of both charged particle track and novel fully automated EM shower reconstruction, a first in a LArTPC.  

We acknowledge the support of the Fermi National Accelerator Laboratory (Fermilab). Fermilab is managed by Fermi Research Alliance, LLC (FRA), acting under Contract No. DE-AC02-07CH11359.  MicroBooNE is supported by the following: the U.S. Department of Energy, Office of Science, Offices of High Energy Physics and Nuclear Physics; the U.S. National Science Foundation; the Swiss National Science Foundation; the Science and Technology Facilities Council of the United Kingdom; and The Royal Society (United Kingdom).  Additional support for the laser calibration system and cosmic ray tagger was provided by the Albert Einstein Center for Fundamental Physics, Bern, Switzerland.

\bibliographystyle{apsrev4-1}
\bibliography{bibliography.bib}
\end{document}